\documentclass[conference,10pt]{IEEEtran}
\IEEEoverridecommandlockouts

\usepackage{cite}
\usepackage{amsmath,amssymb,amsfonts,bm}
\usepackage{algorithm,algorithmic}
\usepackage{graphicx} 
\usepackage{textcomp}

\usepackage{blindtext}
\usepackage{eso-pic}

\makeatletter

\def\ps@IEEEtitlepagestyle{%
	\def\@oddfoot{\mycopyrightnotice}%
	\def\@evenfoot{}%
}
\def\mycopyrightnotice{%
	{\footnotesize  978-1-7281-1508-5/19/ \$31.00 \textcopyright 2019 IEEE\hfill}
	{}
	\gdef\mycopyrightnotice{}
}

\usepackage{eso-pic}
\newcommand\AtPageUpperMyright[2]{\AtPageUpperLeft{%
		\put(\LenToUnit{0.08\paperwidth},\LenToUnit{-1.1cm}){%
			\parbox{0.5\textwidth}{\raggedleft\fontsize{10}{11}\selectfont #1}}%
	}}%
	\newcommand{\conf}[1]{%
		\AddToShipoutPictureBG*{%
			\AtPageUpperMyright{#1}
		}
	}

\begin{document}
\conf{27th Iranian Conference on Electrical Engineering (ICEE2019)}
\title{Spectrum Trading for Device-to-Device Communication In Cellular Networks using Incomplete Information Bandwidth-Auction Game  }

\author{\IEEEauthorblockN{Mohammad Karimzadeh Farshbafan}
\IEEEauthorblockA{{\textit{Dept. of Electrical and Computer Eng.}} \\
\textit{College of Eng., University of Tehran}\\
Tehran, Iran, 14176-14418 \\
m.karimzadeh68@ut.ac.ir}
\and
\IEEEauthorblockN{ Mohammad Hossein Bahonar}
\IEEEauthorblockA{{\textit{Dept. of Electrical and Computer Eng.}} \\
\textit{Isfahan University of Technology}\\
Isfahan, Iran, 84156-83111 \\
mh.bahonar@ec.iut.ac.ir}
\and
\IEEEauthorblockN{Farshid Khajehrayeni}
\IEEEauthorblockA{{\textit{Dept. of Electrical and Computer Eng.}} \\
\textit{Tarbiat Modares University}\\
Tehran, Iran, 14115-111 \\
f.khajehrayeni@modares.ac.ir}
}


\maketitle

\begin{abstract}
Device-to-device (D2D) communication that allows proximity users to communicate directly has been recently proposed to improve spectral efficiency of cellular networks.
In this paper, we assume a cellular network consisting of multiple cellular user equipments (CUEs), which are the primary users, and a cognitive D2D pair, which is the secondary user.
The D2D pair needs a bandwidth for data transmission that can be obtained via spectrum trading.
We introduce a bandwidth-auction game for the spectrum trading problem.
The base station (BS) and CUEs are able to sell their spectrum or share it with the D2D pair, which allows the D2D pair to operate in orthogonal sharing, cellular, or non-orthogonal sharing (NOS) modes.
Operation of the D2D pair in the NOS mode causes interference to the CUEs, which is possible under low interference condition.
In the auction, the D2D pair can buy its required spectrum from three different service providers (SPs) corresponding to each mode that operate on different frequency spectrums.
The D2D pair bids a price-bandwidth demand curve and the SPs offer a price-demand supply curve.
Since each player is not aware of the strategy of other players in practical scenarios, the game is assumed to be an incomplete information repeated one.
A best response based learning method is proposed for the decision making procedure of all players, the D2D pair and SPs.
It is shown that the proposed method converges to the Nash equilibrium (NE) point of the game more rapidly than the state-of-the-art methods when the game is played repeatedly.
The sensitivity of the proposed method to the learning rate variable is also less than the state-of-the-art methods and hence can be considered as a robust one.
\end{abstract}

\IEEEpeerreviewmaketitle
\vspace{-2mm}
\section{Introduction}
Cognitive radio is a promising techniques that has been proposed in order to solve the spectrum scarcity problem \cite{R073}.
In a cognitive radio network, secondary users (SUs) are able to reuse the radio resources of primary users (PUs) while the resources are not being used by PUs or the interference caused to PUs from SUs does not degrade the quality of service of PUs below a predetermined threshold.
In addition, device-to-device (D2D) communication which corresponds to direct data transmission between two proximity user equipments (UEs) without going through the base station (BS) has also been proposed in order to increase the overall spectral efficiency (SE) of cellular networks or to offload part of the cellular traffic under heavy load of cellular networks \cite{R018}.
This communication technique that enables the network to provide service to a larger number of users has been considered in 3GPP LTE-Advanced standard.

Due to the potentials of D2D communication, it has received much attention recently 
\cite{R003,R077,R052,R024_017}.
It is also possible to use D2D communication for cognitive radio systems.
Some researches  and technologies such as \cite{R024_017, R076} are based on the usage of cognitive radio concepts for D2D communication in cellular networks.
D2D communication in the licensed spectrum can be used in an overlay manner where the resources of D2D pairs are different from that of cellular user equipments (CUEs) or an underlay manner where D2D pairs and CUEs use the same set of resources.
The resource allocation problem for D2D communication of D2D pairs has been studied in centralized schemes using optimization techniques \cite{R003} or distributed schemes using game theoretic models \cite{R075}.
Since game theoretic based methods are designed in a distributed manner, they are usually {scalable and less complex} \cite{R010}.
Different types of games such as coalitional game \cite{R012}, dynamic stackelberg game \cite{R013}, and auction-based games \cite{R075} has been used to model and solve the problems that are related to D2D communication in cellular networks.

Auction theory has been employed to model and investigate the spectrum sharing problem in D2D communication \cite{R075} and cognitive radio systems \cite{R078}.
In \cite{R078}, the spectrum access problem of SUs is modeled using a repeated auction game where a Bayesian learning algorithm is proposed as the solution that lead to the Nash equilibrium (NE) point of the game.
Auction games can also be used for spectrum trading or spectrum sharing problem in D2D communication.
In \cite{R079}, a reverse iterative combinatorial auction game has been proposed as the resource allocation method of D2D communication, which leads to a good performance of system sum-rate.
In \cite{R080} a Cournot game is proposed to model the problem of competitive pricing where a few number of firms compete with each other in terms of price in order to gain the highest profit.
However, it is assumed that the price is completely in hands of PUs in this model, and it does not consider the demand curve of the SUs.
In \cite{R081} a double sided auction mechanism is proposed to solve the dynamic spectrum sharing problem.
The game considers the supply and bid strategy for suppliers and bidders, respectively.
Based on the proposed model, the bidders and suppliers do not have any contribution to the amount of bandwidth and price, respectively.
In \cite{R076} the authors propose a bandwidth-auction game for resource allocation to one D2D pair in a cellular network and derive the NE point analytically.
Since the assumption of complete information is not a practical one, a gradient based learning algorithm is also proposed, which converges to the NE point of the game.

In this paper, we assume a cellular network consisting of a base station (BS) and multiple CUEs that are considered as the PUs.
A D2D pair, which is considered as the SU, exists in the cell and needs spectrum resources for data transmission.
The spectrum sharing problem is formulated as a bandwidth-auction game.
The D2D pair can buy its required bandwidth from three types of service providers (SPs) that are willing to share or resale their spectrum.
The BS and CUEs are able to provide bandwidth for the D2D pair.
The D2D pair can use the spectrum of the SPs that are willing to resale their resource in an overlay manner where all UEs use orthogonal resources.
{In this way}, the D2D communication does not cause interference to the other UEs.
The D2D pair is also able to reuse the spectrum of the CUEs in an underlay manner under low interference conditions.
{Therefore}, CUEs share their resource with the D2D pair, and D2D communication causes interference to the other CUEs.
{Hence}, the BS and CUEs are the suppliers of the game while the D2D pair is the consumer and the BS acts as the brooker in the game.
Each palyer aims to maximize its own payoff.
Our system model is mostly based on the system model of \cite{R076}.
The D2D pair bids a demand curve and each supplier offers a supply curve.
In order to have a practical and more realistic system model, it is assumed that each player is not aware of the strategy of the other player and can just be aware of the bandwidth price.
Thus, the game is modeled as a incomplete information game.
In \cite{R076}, it is proposed to use a gradient base learning method for the incomplete information.
We propose to utilize a best response based learning method in a repeated manner as the solution of the incomplete information game.
It is shown that the proposed algorithm converges to the NE point of the game iteratively.
The proposed method converges rapidly and quicker than the state-of-the-art proposed method of \cite{R076}.

The rest of the paper is organized as follows.
In Section \ref{SysModel}, the system model is described and the bandwidth-auction game is presented. 
In Section \ref{ProposedMethod} the NE point of the game is calculated and the learning method for the incomplete game is proposed.
The simulation results are presented in Section \ref{Sec_Simulation}, and Section \ref{Sec_Conclusion} concludes the paper.

\section{System Model}
\label{SysModel}
\subsection{Cellular Network Model}
In this study, a multi cell network is assumed in which each UE can function in either cellular or D2D modes. 
In fact, they communicate with each other through the BS in the cellular mode. 
Moreover, they are allowed to work in a D2D mode on certain conditions. 
First, the UE must be placed in close proximity in order to alleviate the incoming interference to the other CUEs, regardless of being in a cell or not. 
Second, there should be low amount of interference on the receiver of the D2D pair, so that its corresponding SINR remains above a predefined minimum threshold. 
Third, the network must suffer from heavy load and prefer to offload a portion of its traffic load. The UE that satisfies the aforementioned conditions are able to have a D2D communication. 
We call the UEs of this D2D pair as $U_1^d$ and $U_2^d$ which are the transmitter and receiver of the D2D pair, respectively. 
Then, a request is sent to the BS and it allocates the required bandwidth. 

Three various sharing options can be considered including non-orthogonal sharing (NOS), orthogonal sharing (OS), and cellular mode (CM). 
In the case of NOS mode, the same bandwidth with the size of $s_1$, is reused by the D2D pair, resulting in experiencing interferences at the other CUEs caused by the D2D pair and at the receiver side of the D2D pair caused by CUEs. 
In the case of OS mode, one or more CUEs can {resell} their spectrum by providing a portion of their subcarriers to the D2D pair. Hence, there will be no interference caused to the D2D pair by the CUE and also to the CUE caused by the D2D pair.
In this case, the CUEs are denoted by $U_n^c, n=2,...,N+1$ where $N$ is the number of CUEs who are willing to share their spectrum and the corresponding size of the allocated spectrum is denoted by $s_j, j=2,...,N+1$. 
In the case of CM, the operator itself acts as a SP and sells its spare spectrum ($s_{N+2}$). 
Thus, BS performs as a relay node and allocates uplink and downlink bandwidth (denoted by $B_{UL}$ and $B_{DL}$). It should be noted that the communication between users is based on orthogonal frequency division multiple access (OFDMA), which is consistent with LTE advanced network. 
Thus, in each mode, one or more SPs attempt to sell their spectrum to the D2D transmitter at price $\alpha_j^S$.

The game is modeled as a bandwidth-auction game where there exists three kinds of SPs.
The operator, CUEs, and the receiver of the D2D pair in NOS mode act as SPs which are the suppliers of the game.
The D2D pair is the bidder of the game.
Each supplier offers a supply curve and the bidder bids a demand curve.
There are two factors that affect the spectrum demand of the D2D transmitter, namely SE and the price. 
The SE of $U_1^d$ depends on the chosen source (SP). 
In the NOS mode, Let ${\xi _{NOS}}$ be the Signal-to-Interference-plus-Noise-Ratio (SINR) of the D2D pair, $P_{NOS}$ denote the transmit power of $U_1^d$, $BER_{tar}$ be the target bit-error-rate, and $g_{NOS}$ indicate the channel gain of the D2D pair. Then, the SE (throughput) of $U_1^d$ can be written as \cite{R073}:
\vspace{-2mm}
\begin{align} 
T_1 &= \log_2(1+\Theta \xi_{NOS}),~ \rm {where} \label{FirstThroughput}\\ 
\Theta &=\frac{1.5}{\ln(0.2 \times (BER_{tar})^{-1})} , ~\xi_{NOS} = \frac{P_{NOS} g_{NOS} }{\sigma^2 + I} \notag,
\end{align}
in which $\sigma^2$ and $I$ represent the power of additive white Gaussian noise (AWGN) and interference on the receiver of the D2D pair, respectively. 
It must be mentioned that static channels with path-loss model, rayleigh fast fading, and log-normal slow fading are assumed in this mode.
Channel gain between the receiver and transmitter of D2D pair is denoted by $g_{NOS}={(K/d_{1,2})}^p \psi_{1,2} \gamma_{1,2}$ where $K$ is a constant depending on the working frequency, $d_{1,2}$ denotes the distance between the D2D transmitter and receiver, and $p$ (normally between 2 and 6) represents the path-loss parameter, $\psi_{1,2}$ is the rayleigh fast fading gain of the link, and $\gamma_{1,2}$ denotes its corresponding log-normal slow fading gain.

In the OS mode, it is supposed that $\xi_{OS}$ indicates the SNR at the CUE and $P_{OS}$ is the transmit power of $U_1^d$. Then, the throughput of the $U_1^d$ can be expressed as:
\vspace{-2mm}
\begin{align}
T_j &= \log_2{(1+\Theta \xi_{OS})}, \; \rm{where} \label{SecondThroughput}\\
\xi_{OS} &=\frac{P_{OS}g_{OS}}{\sigma^2}, \;\; \text{for} \;\; j=2,3,...,N+1 \notag.
\end{align}

In the CM mode, let $\xi_{CM,1}$ and $\xi_{CM,2}$ be the SNR at the BS and that of at the receiver of the D2D pair, respectively. Furthermore, we denote the transmission power emitted from $U_1^d$ in this mode as $P_{CM}$, the channel gain between the D2D transmitter and the BS as $g_{CM,1}$, and that of between the BS and the D2D receiver as $g_{CM,2}$. 
Then, the throughput of $U_1^d$ can be formulated as \cite{R073}:
\vspace{-2mm}
\begin{equation}
T_{N+2} = \min \left\{
\begin{array}{ll}
\frac{ B_{UL} } { B_{UL}+B_{DL} } \log_2(1+\Theta \xi_{CM,1})\\
\frac{ B_{UL} } { B_{UL}+B_{DL} } {\log_2(1+\Theta \xi_{CM,2}) }
\end{array}					 
\right\},
\label{ThirdThroughput}
\end{equation}
\vspace{-4	mm}
\begin{align}
\nonumber
\xi_{CM,1}= \frac{P_{CM}g_{CM,1}}{\sigma^2},
\xi_{CM,2}= \frac{P_{CM}g_{CM,2}}{\sigma^2}.
\end{align}
If we assume the distance between the D2D pair is long enough, the following formula for the throughput can be derived \cite{R073}:
\vspace{-4mm}
\begin{align}
T_{N+2} &\approx \frac{B_{UL}}{ B_{UL} + B_{DL} } \log_2( 1+ \Theta \xi_{N+2}),\\
\xi_{N+2} &= \frac{ P_{CM} g_{CM} }{ \sigma^2 },
\nonumber 
\end{align}
where $g_{CM}$ is the channel gain between the D2D pair and the BS.

As it is previously mentioned, we assume a situation in which there are $N+2$ SPs ($U_1^d$, $N$ cellular users, and BS) and only one bidder. Each of these players offers a price based on their strategy functions (supply functions for SPs and demand functions for the D2D transmitter), and the players’ desire is to maximize their own payoff. 
This is a uniform price auction game whose aim is to set each player’s strategy in a way that they reach the NE point of the game. 
In other words, both demand and supply functions are known to the broker, which is the BS in our model, and its duty is to calculate the ultimate price by adjusting the bandwidth demand and supply. Having calculated the price, the broker have a duty to allocate the determined bandwidth to the D2D user.
\vspace{-3mm}
\subsection{Bandwidth Acution Game Model}
As it was mentioned before, we consider a D2D pair demanding an amount of bandwidth and $(N+2)$ SPs that compete for providing the requested bandwidth to the D2D pair. 
First of all, a demand function for the D2D pair and supply functions of the SPs are defined.
The considered demand and supply functions for the players of the bandwidth auction game are as follows \cite{R073}:
\vspace{-2mm}
\begin{align}
p &= -\Lambda^D b^D + \lambda^D, \notag \\ 
p &= \Lambda^S_i b^S_i + \lambda^S_i, \;\;\; i = 1,2,\ldots,N+2 \label{SupplyDemandFunction}.
\end{align}
where $b^D$ is the demand bandwidth of the D2D pair, $b^S_i$ is the supply bandwidth of the $i^{\text{th}}$ SP and $p$ is the clearing price for the unit of the bandwidth. 
Also, $\Lambda^D$, $\Lambda^S_i$, $\lambda^D$ and $\lambda^S_i$ are the strategies of different players which are positive numbers. 
The considered functions in \eqref{SupplyDemandFunction} means that D2D user buys more bandwidth at the lower price and vice-versa. 
Also, SPs sell more bandwidth at the higher price and vice-versa. 

The payoff function for the players of the bandwidth auction game must be defined. 
For the D2D player, it can be defined as \cite{R073}:
\vspace{-5.5mm}
\begin{align}
\pi^D= U(b^D) - pb^D, \quad U(b^D) = \sum_{i=1}^{N+2}{R_iT_ib^S_i} \label{DemandSidePayoff},
\end{align}
where $U(b^D)$ is the utility function of the D2D player, $R_i$ is its revenue when using the bandwidth of the $i^{\text{th}}$ SP and $T_i$ is the throughput of the D2D pair when using the bandwidth of the $i^{\text{th}}$ SP which is introduced in \eqref{FirstThroughput}-\eqref{ThirdThroughput}.

The payoff function for the SPs can also be written as 
\begin{align}
\pi^S_i=  pb^S_i - C_i(b^S_i), \quad C_i(b^S_i) = c_ib^S_i \label{SupplySidePayoff},
\end{align}
where $C_i(b^S_i)$ is the cost of providing bandwidth for the $i^{\text{th}}$ SP. 
As seen in \eqref{SupplySidePayoff}, the $C_i(b^S_i)$ for all of the SPs is considered linearly proportional to the sold bandwidth, $b^S_i$. 
However, the unit cost of providing bandwidth, $c_i$, is determined differently for the SPs. 
More precisely, for the first SP, the unit cost of providing bandwidth can be written as $c_1 = c_1^{BS}+c_1^I$ in which $c_1^{BS}$ is the unit price of the bandwidth that first SP has paid to the BS and $c_1^I$ is the unit cost of the imposed interference to the first SP. 
As mentioned earlier, the second to the $(N+1)^{\text{th}}$ SPs are the cellular users. As a result, the $c_i, \; i=2,3,\ldots,N+1$ is the price of purchasing bandwidth from the BS for the $(i-1)^{\text{th}}$ cellular user. Finally, for the last SP which is the BS, $c_{N+2}$ is the price of providing bandwidth.
\vspace{-2mm}
\section{Proposed Method for Bandwidth Auction Game}
\label{ProposedMethod}
In this section the proposed method is described.
We propose a learning method for the incomplete information model of the game which is more reasonable and practical in section \ref{SecIncomplete}.
In order to evaluate the performance of the proposed method, the NE point of the complete information game is also calculated in section \ref{SecComplete}.
\vspace{-2mm}
\subsection{Complete Information Game}
\label{SecComplete}
In a complete information game, all the players are aware of the strategies, payoffs, and demand/supply functions of the other players.
Using the demand and supply functions in \eqref{SupplyDemandFunction} and also the fact that $b^D = \sum_{i=1}^{N+2}{b_i^S}$, the clearing price can be written as 
\vspace{-5mm}
\begin{align}
p = \dfrac {\dfrac{\lambda^D}{\Lambda^D}+\sum_{i=1}^{N+2}{\dfrac{\lambda^S_i}{\Lambda^S_i}}}{\dfrac{1}{\Lambda^D}+\sum_{i=1}^{N+2}{\dfrac{1}{\Lambda^S_i}}}.
\label{clearingPrice}
\end{align}
When the price is cleared using \eqref{clearingPrice}, each player can determine the amount of purchased or sold bandwidth using \eqref{SupplyDemandFunction}.

It is known that all of the players are willing to maximize their payoff function which is a function of the clearing price. The players can maximize their payoff by adjusting their strategy $\big( \Lambda^D, \lambda^D, \Lambda^S_i, \lambda^S_i\big)$. 
On the other hand, the unit price of bandwidth, $p$ is a function of all players' strategies. 
As a result, the strategy of each player will affect the payoff of the other players. 
The problem with this characteristic can be modeled as a game.

As mentioned earlier, the strategy of each player is to adjust the coefficients of the demand/supply functions which are introduced in \eqref{SupplyDemandFunction}. 
However, when both of the $\Lambda^D, \lambda^D$ for the D2D user and $\Lambda^S_i, \lambda^S_i$ for the SPs are considered as the strategies of the players, the mathematical approach of NE calculation becomes intractable. 
As a result, only one of the coefficients on the demand/supply functions should be selected as the strategy of the players. 
The authors in \cite{R075} proved that if the $\Lambda^D$ for the D2D user and $\Lambda^S_i$ for the SPs are selected as the strategy of the players and $\lambda^D$ and $\lambda^S_i$ are considered constant and equal to their optimal values, the NE point does not exist. 
Hence, $\lambda^D$ and $\lambda^S_i$ are considered as the strategy of the players. Moreover, it is assumed that the $\Lambda^D$ and $\Lambda^S_i$ are positive and have constant values. 

{Therefore}, the NE point of the game while  $\lambda^D$ and $\lambda^S_i$ are considered as the strategy of the players and $\Lambda^D$ and $\Lambda^S_i$ as positive constants can be calculated. 
The NE point is the intersection point of the best response of all players. Considering this method and based on the solution presented at \cite{R075}, the NE point can be calculated as follows
\begin{align}
\bm{ \mathrm{ \lambda } } ^* = \bm{ \mathrm{ A }}^{-1} \bm{ \mathrm{ B } } \label{NE_Formula},
\end{align}
in which $ \bm{ \mathrm{ \lambda } }=\left[ {\lambda_1^s, \ldots, \lambda_{N+2}^s, \lambda^D} \right]^T \in \Re^{N+3}$ is the ultimate price related to the NE point. ordered lexicographically as a column vector, $\bm{ \mathrm{ A }}= \left[{ a_{i,j} } \right] \in \Re^{(N+3) \times (N+3) }$ is an invertible  square matrix ($det(\bm{ \mathrm{ A } }) \ne 0 $), and $ \bm{ \mathrm{ B } } = \left[ { B_1, \ldots, B_{N+2}, B_{N+3} } \right]^T \in \Re^{N+3} $ is a column vector. The following equation illustrates how elements of the mentioned matrices are calculated:
\vspace{-2mm}
\begin{align}
	a_{i,i} &= 2 ( \Phi \Lambda_i^s )^{-1} - 2, 
	\nonumber \\
	a_{i,j} &= 2 ( \Phi \Lambda_j^s )^{-1} - \Lambda_i^s (\Lambda_j^s)^{-1}, \quad  i \ne j,
	\nonumber \\
	a_{i,{N+3}} &= 2(\Phi \Lambda^D)^{-1} - \Lambda_i^s (\Lambda^D)^{-1},
	\nonumber \\
	a_{{N+3},j} &= 2(\Phi \Lambda_j^s)^{-1} - \Lambda^D,
	\nonumber \\
	a_{{N+3},{N+3} } &= 2 (\Phi \Lambda^D)^{-1} - 2,
	\nonumber \\
	B_i &= c_i (1-\Phi \Lambda_i^s), \quad \Phi = \sum\limits_{n=1}^{N+2} ({\Lambda_n^s})^{-1} + (\Lambda^D)^{-1},
	\nonumber	\\
	B_{N+3} &= -\Lambda^D \sum\limits_{n=1}^{N+2} R_n T_n (\Lambda_n^s)^{-1},
	\nonumber\\
	i,j &= 1,2,\ldots,N+2.
	\nonumber	
\end{align}

It should be noted that if matrix $\bm{ \mathrm{ A} } $ is not an invertible matrix ($det( \bm{ \mathrm{A} } = 0 $), there will be no NE point defined for the game.

The most important assumption of using \eqref{NE_Formula} for calculating the NE point is the complete information assumption of the game.
More precisely, for calculation of the NE point using \eqref{NE_Formula}, each player should be informed about the payoff function and demand/supply functions of the other players which is not possible in a practical scenario. 

\subsection{Incomplete Information Game}
\label{SecIncomplete}
The previous sections derive the NE point of the game considering the complete information assumption which is not a reasonable assumption in a practical scenario. Practically, each player cannot be aware of the payoff, demand/supply, or strategy of other players.
In fact, the bandwidth information is the only information each player has. 
According to these explanations, the players only can reach the NE point of the game in multiple steps. 
More precisely, they should learn to improve their strategy until reaching the NE point. 
In \cite{R075}, a learning method for adjusting the strategy of the players based on the marginal profit function is introduced. In this method, the players make their strategy in each step based on a local estimation of their marginal payoffs. This adjustment mechanism is called myopic.
We propose the best response based learning method for the incomplete information game model.

In the best response based learning, each player utilizes the best response function for updating the playing strategy in each step. As a result, the strategy of each player of the considered game can be written as follows
\begin{align}
\lambda^D(t+1)&=f^D\big(\lambda^S_1(t), \ldots, \lambda^S_{N+2}(t)\big), \label{update1}\\
\lambda^S_i(t+1)&= \label{update2}\\
f_{i}^{S}\big(\lambda^D(t), &\lambda^S_1(t), \ldots, \lambda^S_{i-1}(t), \lambda^S_{i+1}(t), \ldots,\lambda^S_{N+2}(t)\big) \notag,
\end{align}
where $f^D$ and $f_{i}^S$ are the best response function of the D2D user and the $i^{\text{th}}$ SP, respectively. This updating approach might lead to large fluctuations in the strategy of each player due to not using the strategy of the player in the previous step. As a result, converging to the Nash equilibrium point cannot be guaranteed. Owing to these challenges, the updating formula in \eqref{update1}-\eqref{update2} are revised as follows
\begin{align}
\lambda^D(t+1)&=(1-\tau^D)\lambda^D(t) + \tau^Df^D\big(\lambda^S_1(t), \ldots, \lambda^S_{N+2}(t)\big) \label{update12}\\
\lambda^S_i(t+1)&= (1-\tau^S_i)\lambda^S_i(t) + \label{update22}\\
\tau^S_if_{i}^{S}\big(\lambda^D&(t), \lambda^S_1(t), \ldots, \lambda^S_{i-1}(t), \lambda^S_{i+1}(t), \ldots,\lambda^S_{N+2}(t)\big) \notag,
\end{align}
where $0 < \tau^D \leq 1$ and $0 < \tau^S_i \leq 1$ are the learning rates of the D2D user and the $i^{\text{th}}$ SP, respectively. For determining the best response function for the D2D user and SPs, the clearing price of the last step is used. As a result, the updating formula in \eqref{update12}-\eqref{update22} can be written as
\begin{align}
\lambda^D(t+1)=&(1-\tau^D)\lambda^D(t)+ \label{update13}\\
& \tau^D\Big(p(t)(2-K(t)\Lambda^D) + \Lambda^D C(t)\Big), \notag\\
\lambda^S_i(t+1)=& (1-\tau^S_i)\lambda^S_i(t) + \label{update23}\\
&\tau^S_i\Big(p(t)+(p(t)-\Lambda_i^S)(1-K(t)\Lambda_i^S)\Big) \notag,
\end{align}
where $p(t)$ is the last clearing price. Also, $C(t)$ and $K(t)$, which are unknown parameters to the players, can formulated as
\vspace{-4mm}
\begin{align}
K(t) \simeq & \dfrac{2p(t)+\Lambda^DC(t)}{p(t)\Lambda^D+(\Lambda^D)^2\phi^D(t,t-1)} \label{K_Formula}\\
\phi^D(t,t-1) =& \dfrac{\pi_D(t)-\pi_D(t-1)}{\Lambda_D(t)-\Lambda_D(t-1)}\\
C(t)=&\sum_{i=1}^{N+2}{\dfrac{R_iT_i}{\Lambda_i^S}}
\end{align}

As it can be seen, both of these  unknown parameters depend on the strategies of all players. 
As a result, each player should estimate them through the learning procedure. 
However, as described in \eqref{K_Formula}, $K(t)$ is a function of $C(t)$. 
Hence, it is assumed that the value of $K(t)$ is updated in {odd time step} using \eqref{K_Formula} and $C(t) = C(t-1)$. 
Also, the value of $C(t)$ is updated {in even time step} using \eqref{C_Formula} and $K(t) =K(t-1)$.
\begin{align}
C(t)=&\dfrac{K(t)\Big(p(t)\Lambda^D+(\Lambda^D)^2\phi^D(t,t-1) - 2p(t)\Big)}{\Lambda^D} \label{C_Formula}
\end{align}

\section{Simulation Results}
\label{Sec_Simulation}
In this section, we evaluate the performance of the proposed learning method using numerical results.
We consider a single cell cognitive radio-based cellular system with one BS, one D2D pair, and one CUE where the BS is placed at the middle point of the cell.
CUEs are uniformly distributed in the cell using 2D uniform distribution.
Due to the short distance between transmitter and receiver of a D2D pair, the transmitter of D2D pair is uniformly distributed in the cell while the receiver of the D2D pair is placed uniformly in a cluster around the transmitter.

To draw a comparison among state-of-the-art methods in the incomplete information case, we employed the presented simulation setup in \cite{R075} that used a gradient-based learning method. 
The parameters of SPs and the D2D pair are presented in \eqref{EqPar1}.
\vspace{-2mm}
\begin{align}
\label{EqPar1}
\Lambda^S_1 = 0.6, \Lambda^S_2 = 0.45,  ~\Lambda^S_3 = 0.6, \rm{and} \Lambda^D = 0.5
\end{align}
Also the target BER for the users is assumed to be $BER_{tar} = 10^{-4}$.
Furthermore, the channel qualities of these SPs in NOS mode, OS mode, and CM which are  represented in terms of SINR are introduced in \eqref{EqPar3}.
\vspace{-2mm}
\begin{equation}
\label{EqPar3}
\xi_{NOS}=20\rm{dB},~ \xi_{OS}=30 \rm{dB}, ~\rm{and}~ \xi_{CM}=10 \rm{dB}
\end{equation}
The channel quality of CM is lower than that of other modes due to the large distance between a user and the BS.
The channel quality of NOS mode is also lower than that of OS mode due to low amount of interference.
Considering the described simulation parameters, there can be a fair comparison between the our simulation results and those of \cite{R075}.

Simulation results verify the convergence of the players' strategies to the NE point of the game after sufficient number of time steps.
After a sufficient number of time steps, the strategies of $U_2^d$, $U^c$, BS, and the D2D user converge to $(4.8656, 4.3152, 7.6923, 24.9058)$.
By calculating the NE point of the game in the complete information case using equation \eqref{NE_Formula}, it can be shown that the NE point of the game $\bm{ \mathrm{ \lambda } } ^*$ is equal to $\bm{ \mathrm{ \lambda } } ^* = (4.8656, 4.3152, 7.6923, 24.9058)$.
Hence the best response based proposed method converges to the NE point of the game.

\begin{figure}[!t]
\includegraphics[scale=0.42]{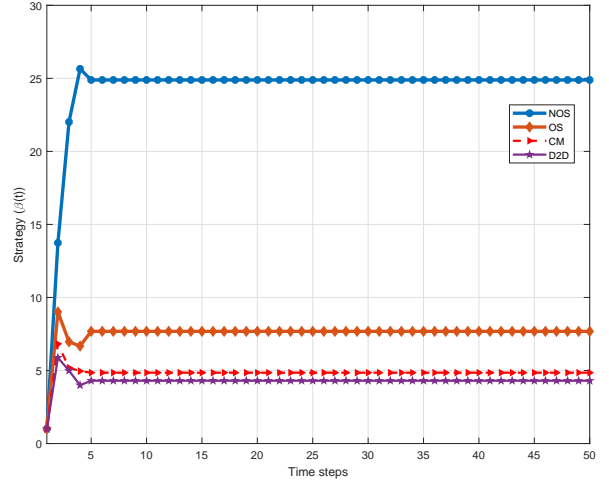}
\caption{\small{Strategies of the SPs and D2D user during the learning
game and convergence to the NE point when learning rates are set to $0.6$.}}
\label{Fig1}
\end{figure}

In the following, we evaluate the performance of two learning method including gradient-based learning and the proposed best response based learning regarding their convergence to the NE point in the incomplete information case. 
In Fig. \ref{Fig1}, the strategies of the SPs and the D2D user during the learning game, using best response method and convergence to the NE point are indicated. 
The learning rates of all players are set to $0.6$. As seen in this figure, the strategies of all players converge to the NEm point of complete information case. It is worth noting that the convergence occurs only in $5$ time steps. 
However, as indicated in \cite{R075}, the convergence of strategies using gradient-based learning method occurs in $9$ time steps. As a result, the convergence speed of the proposed algorithm is higher than gradient-based learning method.

\begin{figure}[!t]
\includegraphics[scale=0.41]{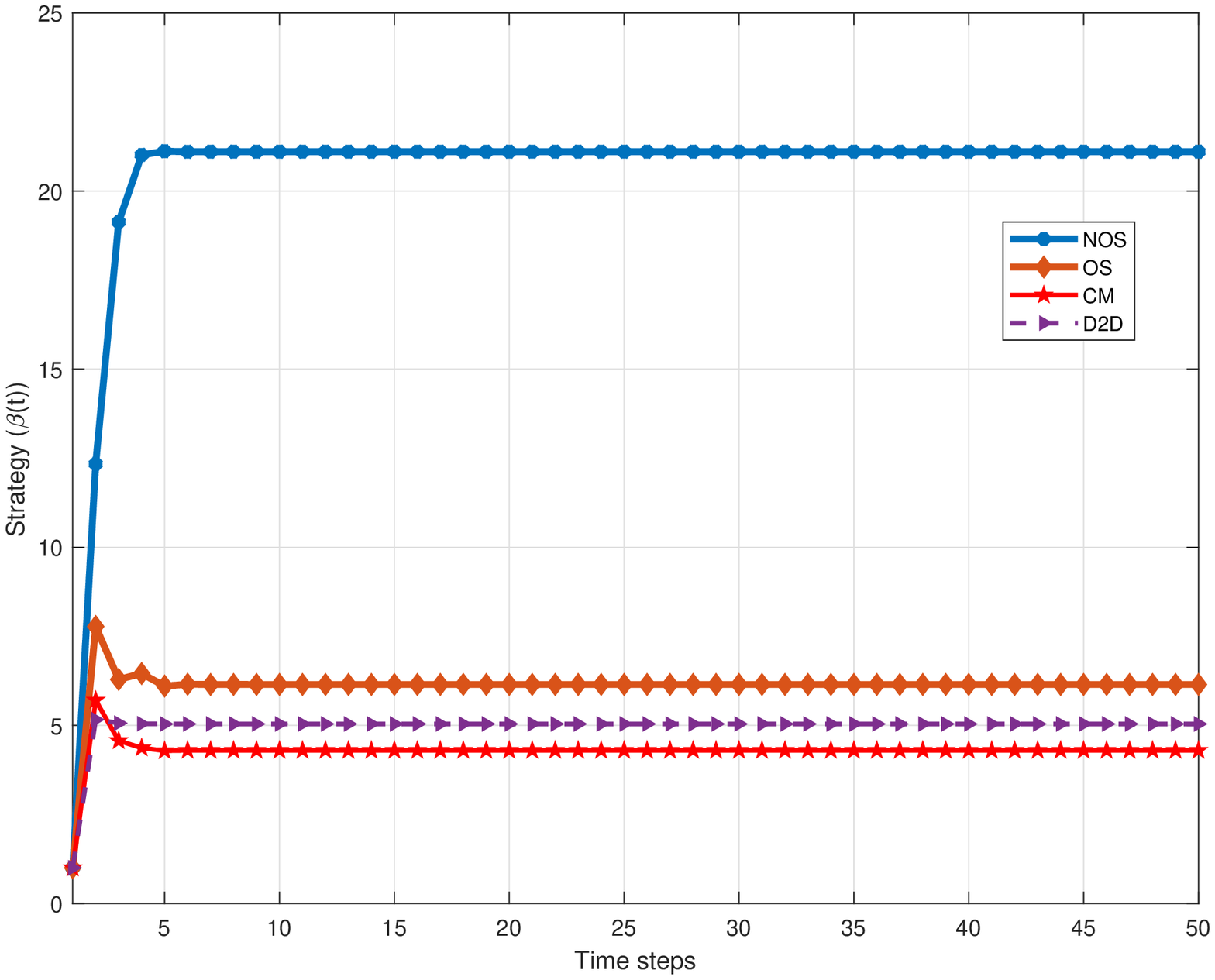}
\caption{\small{Strategies of the SPs and D2D user during the learning
game and convergence to the NE point of the game when learning rates of the players are set to $0.4$.}}
\label{Fig2}
\end{figure}

\begin{figure}[!t]
\includegraphics[scale=0.41]{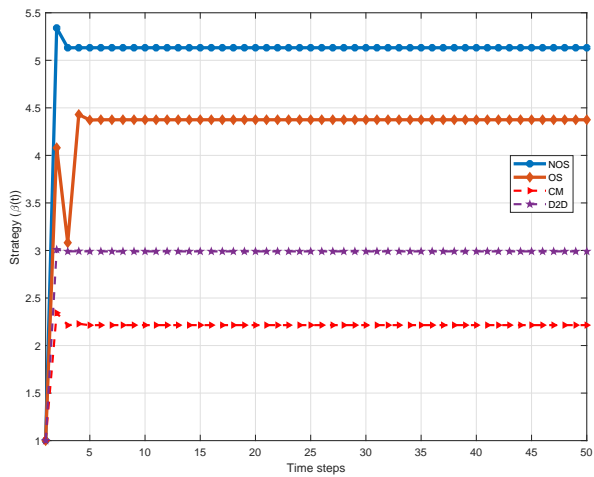}
\caption{\small{Strategies of the SPs and D2D user during the learning
game and convergence to the NE point of the game when learning rates of the players are set to $0.1$.}}
\label{Fig3}
\end{figure}

The effect of different learning rates on the convergence of the players' strategy is considered in Fig. \ref{Fig2} and Fig. \ref{Fig3}. 
More precisely, Fig. \ref{Fig2} and Fig. \ref{Fig3} demonstrate the strategy of the players for different time steps of the learning game when the learning rate is set to $0.4$ and $0.1$, respectively. 
As it can be seen in these figures, decrease in the learning rate value causes the strategy of the players to get away from the NE point of the game. 
However, the convergence speed of the proposed method has not changed. 
It is worth noting that the strategy of the players in gradient-based learning is very sensitive to the learning rate, as indicated in \cite{R075}.
So it can be concluded that the proposed method is more robust compared to the gradient-based learning method.

\vspace{-3mm}
\section{Conclusion}
\label{Sec_Conclusion}
In this study, we have simulated a spectrum trading problem in a cellular network with primary users and a secondary user (a cognitive D2D pair). 
In the considered problem, the primary users can share their spectrum with the D2D pair.
They are also able to resell their bandwidth to the D2D pair.
Since each user tries to optimize its own benefit, the problem can be modeled by a bandwidth-auction game.
Owing to being practically operative, the game has been assumed to be an incomplete information repeated one as no user is aware of the others' strategy in a practical scenario. 
We have addressed the problem by our proposed best response based learning approach. 
The NE point of the game in an complete information scenario is derived and it has been demonstrated that the proposed method converges to the NE point of the game in an iterative manner. 
The simulation results have confirmed that the proposed approach outperforms the state-of-the-art methods in terms of convergence speed. 
It has been shown that the the proposed learning method can be considered as a robust one regarding the learning rate when it is compared with the state-of-the-art methods.
Thus, the proposed best response learning based method is an effective approach in terms of performance and convergence speed.
\vspace{-2mm}
\bibliographystyle{IEEEtran}
\bibliography{Test_GameRes}

\end{document}